\documentstyle[prl,aps,epsf,multicol]{revtex}

\begin{document}
\draft

\title{ Spin Order due to Orbital Fluctuations: Cubic Vanadates }

\author { Giniyat Khaliullin,$^1$ Peter Horsch,$^1$ 
          and Andrzej M. Ole\'{s}$^{1,2}$ }
\address{ $^1$Max-Planck-Institut f\"ur Festk\"orperforschung,
              Heisenbergstrasse 1, D-70569 Stuttgart, Germany }
\address{ $^2$Institute of Physics, Jagellonian University, 
              Reymonta 4, PL-30059 Krak\'ow, Poland }

\date{March 14, 2001}
\maketitle

\begin{abstract}
We investigate the highly frustrated spin and orbital superexchange 
interactions in cubic vanadates. The fluctuations of $t_{2g}$ orbitals 
trigger a {\it novel mechanism of ferromagnetic interactions\/} between 
spins $S=1$ of V$^{3+}$ ions along one of the cubic directions which 
operates already in the absence of Hund's rule exchange $J_H$, and 
leads to the C-type antiferromagnetic phase in LaVO$_3$. The Jahn-Teller 
effect can stabilize the orbital ordering and the G-type antiferromagnetic 
phase at low temperatures, but large entropy due to orbital fluctuations 
favors again the C-phase at higher temperatures, as observed in YVO$_3$. 
\end{abstract}

\pacs{PACS numbers: 75.30.Vn, 71.27.+a, 75.30.Et, 79.60.-i.}

\begin{multicols}{2} 
Large Coulomb interactions play a crucial role in transition metal oxides, 
and are responsible for the collective behavior of strongly correlated $d$ 
electrons which localize in Mott-Hubbard (or charge-transfer) insulators 
\cite{Ima98}. Such localized electrons may occupy degenerate orbital states 
which makes it necessary to consider orbital degrees of freedom at equal
footing with electron spins, and leads to the effective (superexchange) 
spin-orbital models to describe the low-energy physics 
\cite{Kug82,Tok00,Ole00}. A remarkable feature of these models is that the
superexchange interaction is {\it highly frustrated\/} on a cubic lattice, 
which was recognized as the origin of novel quantum effects in transition 
metal oxides \cite{Fei97}. In case of $e_g$ orbital systems this frustration 
is likely removed by orbital order due to order-out-of-disorder mechanism, 
which maximizes the energy gain from quantum spin fluctuations \cite{Kha97}.   
Moreover, quantum effects among $e_g$ orbitals are largely suppressed by the 
Jahn-Teller (JT) effect in real systems, which together with superexchange 
often leads to structural phase transitions accompanied by a certain ordering 
of occupied orbitals, supporting particular magnetic structures. Some well 
known examples are systems with degenerate $e_g$ orbitals filled either 
by one hole (KCuF$_3$), or by one electron (LaMnO$_3$), which order
antiferromagnetically well below the structural transition. 

The transition metal oxides with partly filled $t_{2g}$ orbitals exhibit 
different and more interesting phenomena. This occurs due to the relative 
weakness of the JT coupling in this case, and due to the higher degeneracy 
and additional symmetry of $t_{2g}$ orbitals \cite{Kha00}. As a result, the 
orbitals may form the {\it coherent orbital-liquid\/} ground state stabilized 
by quantum effects, as observed in the spin $S=1/2$ Mott-insulator LaTiO$_3$ 
\cite{Kei00}. It is puzzling what happens when the $t_{2g}$ orbitals are 
filled by two electrons, as in vanadium oxides. On one hand, the occupied 
$t_{2g}$ orbitals are known to order in non-cubic vanadium compounds, such 
as LiVO$_2$ \cite{Pen97} and V$_2$O$_3$ \cite{Mil00}. In fact, the first 
spin-orbital model for V$_2$O$_3$ with spins $s=1/2$ was proposed over twenty 
years ago \cite{Cas78}, but later it was realized that $J_H$ at V$^{3+}(d^2)$ 
ions is large \cite{Miz96}, and the relevant model has to involve $S=1$ spins 
\cite{Mil00}. On the other hand, the situation in cubic systems might be 
very different as all the bonds are {\it a priori\/} magnetically equivalent, 
and quantum fluctuations among orbitals are expected to play an important 
role in this case. 
 
In this Letter we derive and investigate the spin-orbital model for cubic 
vanadates: LaVO$_3$ and YVO$_3$. The magnetic order in LaVO$_3$ is C-type
[ferromagnetic chains along $c$-axis which stagger within $(a,b)$ planes], 
with the N\'eel temperature $T_N=140$ K \cite{Mah92,Goo95}, while it is 
staggered in all three directions (G-type) in YVO$_3$ at $T<77$ K and C-type 
at higher temperatures $77<T<114$ K \cite{Goo95,Kaw94,Ren00}. The C-phase is 
particularly surprising as arising from a practically undistorted structure 
of LaVO$_3$ at $T>T_N$ \cite{Goo95}. Recent Hartree-Fock studies have shown 
that indeed C- and G-phase are energetically close \cite{Miz99}. In order to 
understand the microscopic origin of their competition we consider the 
regime of large $U$, and address below the following questions:
  (i) can the superexchange interactions {\it alone\/} explain why the 
      ferromagnetic (FM) and antiferromagnetic (AF) interactions coexist 
      in LaVO$_3$ in spite of a practically ideal cubic structure at $T>T_N$ 
      with almost equal V--V bonds; 
 (ii) why the structural transition in LaVO$_3$ occurs only {\it below\/}
      the magnetic transition; and   
(iii) why is the G-type AF order stable in the low-temperature phase of
      YVO$_3$, while the C-type order {\it wins\/} at higher temperatures?    

We start with a Mott-insulator picture of cubic vanadites, consistent with
the large value of on-site intraorbital Coulomb element $U\simeq 4.5$ eV 
\cite{Miz96}, and with the results of electronic-structure calculations 
\cite{Saw98}. Due to the Hund's coupling $J_H\simeq 0.68$ eV \cite{Miz96} 
the V$^{3+}$ ions are in triplet configuration $^3T_2$. Each $t_{2g}$ orbital 
is orthogonal to one cubic axis. For instance, $yz$ is orthogonal to $a$ axis 
and will be labelled as $a$, while $zx$ and $xy$ will be labelled as $b$ and 
$c$, respectively. The electron densities at V$^{3+}$ ions satisfy a local 
constraint, $n_{ia}+n_{ib}+n_{ic}=2$.

The superexchange interactions between $S=1$ spins arise from the virtual 
excitations $d^2_id^2_j\rightarrow d^3_id^1_j$ on a given bond 
$\langle ij\rangle$, with the hopping $t$ allowed only between {\it two 
out of three\/} $t_{2g}$ orbitals. The $d^3_i$ excited state may be either
a high-spin $^4A_2$ state, or one of three low-spin states: $^2E$, $^2T_1$  
or $^2T_2$ \cite{notex}. When the second order processes 
$d^2_id^2_j\rightarrow d^3_id^1_j\rightarrow d^2_id^2_j$ are analyzed, 
one has to project the $d^3_i$ ($d^2_i$) configuration generated after an 
individual hopping process on the above $d^3_i$ eigenstates ($^3T_2$ ground
state). This leads to the spin-orbital Hamiltonian,   
\begin{equation}
{\cal H}=J\sum_{\gamma}\sum_{\langle ij\rangle\parallel\gamma}\left[ 
    ({\vec S}_i\cdot {\vec S}_j+1)
    {\hat J}_{ij}^{(\gamma)} + {\hat K}_{ij}^{(\gamma)}\right],
\label{model}
\end{equation}
where the orbital operators ${\hat J}_{ij}^{(\gamma)}$ and
${\hat K}_{ij}^{(\gamma)}$ follow from the processes active on the bond 
$\langle ij\rangle\parallel\gamma$, where $\gamma=a,b,c$:
\begin{eqnarray}
\label{orbj}
{\hat J}_{ij}^{(\gamma)}&=&
\case{1}{2}\left[(1+2\eta R)
\left({\vec\tau}_i\cdot {\vec\tau}_j
     +\case{1}{4}n_i^{}n_j^{}\right)\right.       \nonumber \\ 
&-& \left. 
 \eta r\left(\tau_i^z\tau_j^z+\case{1}{4}n_i^{}n_j^{}\right)
-\case{1}{2}\eta R(n_i+n_j)\right]^{(\gamma)},                         \\ 
\label{orbk}
{\hat K}_{ij}^{(\gamma)}&=&
\left[\eta R
\left({\vec\tau}_i\cdot {\vec\tau}_j+\case{1}{4}n_i^{}n_j^{}\right)  
 +\eta r\left(\tau_i^z\tau_j^z
             +\case{1}{4}n_i^{}n_j^{}\right)\right.          \nonumber \\ 
&-&\left. 
   \case{1}{4}(1+\eta R)(n_i+n_j)\right]^{(\gamma)},
\end{eqnarray}
and $J=4t^2/U$. The coefficients $R=1/(1-3\eta)$ and $r=1/(1+2\eta)$ 
originate from the multiplet structure of the $t_{2g}^3$ excited states via 
$\eta=J_H/U$ \cite{notex}. The operators 
${\vec\tau}_i=\{\tau_i^x,\tau_i^y,\tau_i^z\}$ are defined in the orbital 
pseudospin subspace spanned by two orbital flavours which are 
{\it active along a given direction\/} $\gamma$. For instance, for a bond 
$\langle ij\rangle\parallel c$, the interactions follow from the electron 
hopping between the pairs of $a$ and $b$ orbitals, and may be expressed by 
Schwinger bosons:
$\tau_i^+=a_i^{\dagger}b_i^{}$, $\tau_i^-=b_i^{\dagger}a_i^{}$, 
$\tau_i^z=\case{1}{2}(n_{ia}^{}-n_{ib}^{})$, and 
$n_i^{(c)}=n_{ia}^{}+n_{ib}^{}$, and therefore:  
\begin{eqnarray}
\label{heis}
2\left({\vec\tau}_i\cdot {\vec\tau}_j\!+\case{1}{4}n_i^{}n_j^{}\right)^{(c)}&=&
(n_{ia}^{}n_{ja}^{}+a_i^{\dagger}b_i^{}b_j^{\dagger}a_j^{})+
(a \leftrightarrow b), \nonumber \\ 
2\left(\tau_i^z\tau_j^z+\case{1}{4}n_i^{}n_j^{}\right)^{(c)}&=&
 n_{ia}^{}n_{ja}^{}+n_{ib}^{}n_{jb}^{}.
\label{nn}
\end{eqnarray}

Consider first the interactions in the $J_H\to 0$ limit:
\begin{equation}
{\cal H}_0=\case{1}{2}J\sum_{\gamma}\sum_{\langle ij\rangle\parallel\gamma}
({\vec S}_i\cdot {\vec S}_j+1)
\left({\vec\tau}_i\cdot {\vec\tau}_j+\case{1}{4}n_i^{}n_j^{}\right)^{(\gamma)},
\label{pauli}
\end{equation}
where a constant energy of $-2J$ per V$^{3+}$ ion is neglected. It is 
straighforward to understand why the interactions in this limit turn out 
to have the same structure as in LaTiO$_3$ \cite{Kha00}, where for $s=1/2$ 
spins of Ti$^{3+}$ ions one finds instead 
$({\vec s}_i\cdot {\vec s}_j+\case{1}{4})$. In fact, the spin interactions 
follow entirely from the {\it Pauli principle\/}, as the terms 
$\propto {\vec S}_i\cdot {\vec S}_j$ due to the high-spin $^4A_2$ and 
low-spin $^2E$ states, which involve $d^3\{abc\}$ configurations, cancel 
each other.

A remarkable feature of the $t_{2g}$ superexchange in Eq. (\ref{pauli}) is 
that every bond is represented by two equivalent orbitals giving a SU(2) 
symmetric structure
$({\vec\tau}_i\cdot {\vec\tau}_j+\case{1}{4}n_i^{}n_j^{})^{(\gamma)}$
of the orbital part. Depending on the type of orbital correlations this may 
result in a spin coupling constant of either sign. This important property 
resembles that of one-dimensional (1D) SU(4) model \cite{Li98}. The present 
problem is however more involved since there are {\it three $t_{2g}$ 
flavours\/} in a cubic crystal, and SU(2) orbital correlations among two of 
them along a particular direction will necessarily frustrate those 
correlations in other directions. One may also notice a certain analogy with 
the models of valence bond solids \cite{Aff87}: Actually, a large orbital 
moment $L=1$ of $t_{2g}$ states is formally decomposed in Eq. (\ref{pauli}) 
onto pseudospins one-half, active on different bonds. The analogy is again 
only partial since independent rotations within orbital doublets active on 
different bonds are not allowed here by construction, and thus the formation 
of orbital singlets in all directions simultaneously is impossible. 

Another key observation is the difference between the interactions 
derived for the pairs of Ti$^{3+}$ and V$^{3+}$ ions: 
In the $d^1$ configuration spin $s=1/2$ is small, and the idea of composite 
spin-orbital resonance, in analogy to SU(4) excitations \cite{Li98}, 
helped to resolve the orbital frustration problem \cite{Kha00}. In that 
case the superexchange is best optimized by the spin-orbital resonance in all 
three directions, and the orbitals form a three-dimensional quantum liquid 
which coexists with weak spin order of G-type.
This mechanism is however suppressed in the present case of large 
spin $S=1$ at $d^2$ ions, and the quantum energy can be gained mainly from 
the orbital sector. This suggests that a particular {\it classical spin 
configuration\/} may be picked up which maximizes the energy gain from 
{\it orbital fluctuations\/}. Indeed, orbital singlets 
(with $n_{ia}+n_{ib}=1$) may form on the bonds parallel to $c$ axis, 
exploiting fully the SU(2) symmetry of the orbital interactions in one 
direction, while the second electron occupies the third $t_{2g}$ orbital 
($n_{ic}=1$), controlling spin interactions in the $(a,b)$ planes.
  
In order to understand why orbital fluctuations support the C-AF type spin 
order, it is instructive to start with a single bond along $c$-axis. A 
crucial observation is that the lowest energy of $-J/2$ is obtained when the 
spins are {\it ferromagnetic\/}, and the orbitals $a$ and $b$ form a singlet, 
with $\langle {\vec\tau}_i\cdot {\vec\tau}_j\rangle^{(c)}=-\case{3}{4}$ 
\cite{notet}. Thus, one finds a {\it novel mechanism of ferromagnetic 
interactions\/} which operates due to local fluctuations of $a$ and $b$ 
orbitals. At the same time, the orbital resonance on the bonds in $(a,b)$ 
planes is blocked, as $n_{ic}=n_{jc}=1$, and the superexchange is AF due to 
the excitations to $^2T_1$ and $^2T_2$ states with doubly occupied $c$ 
orbitals. Such an electron distribution and the formation  of quasi-1D 
orbital pseudospin chains supports FM spin order along $c$ axis in C-phase, 
and could be stable only at low temperature when a coherent spin state 
breaking the cubic symmetry is formed as well. The onset of the magnetic 
order which coexists with such orbital fluctuations explains also why a 
structural transition is here induced by this coherent electronic state. 
         
\begin{figure}[h]
      \epsfxsize=60mm
      \centerline{\epsffile{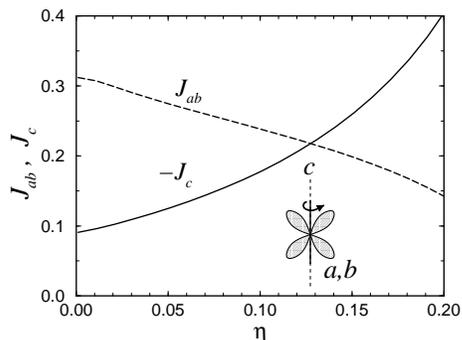}}
\vspace*{0mm}
\caption{
Exchange interactions $J_c$ and $J_{ab}$ (\protect\ref{jcaf}) in C-phase 
(in units of $J$) as functions of $\eta=J_H/U$. The inset indicates the 
local fluctuation between $a\;(=yz)$ and $b\;(=zx)$ orbitals due to singlet 
formation along the $c$ axis. } 
\label{fig:caf}
\end{figure}

We compared the ground state energies of C- and G-phase using the spin-wave
theory for the spin part, while either the exact Bethe ansatz result, or the 
Gaussian fluctuations around the ordered state, were used for the orbital 
part. The exchange constants within $(a,b)$ planes ($J_{ab}>0$) and along 
$c$ axis ($J_c<0$) determine the spin waves. They follow from Eqs. 
(\ref{model})--(\ref{orbk}) (in units of $J$):
\begin{eqnarray}
\label{jcaf}
J_c&=&\case{1}{2}\left[(1\!+\!2\eta R)
 \left\langle{\vec\tau}_i\cdot {\vec\tau}_j\!+\!\case{1}{4}\right\rangle^{(c)}
\!\!-\!\eta r\left\langle\tau_i^z\tau_j^z\!+\!\case{1}{4}\right\rangle^{(c)}
\!\!-\!\eta R\right],                                    \nonumber \\   
J_{ab}&=&\case{1}{4}\left[1-\eta (R\!+\!r)
 +(1+2\eta R-\eta r)\langle n_{ia}n_{ja}\rangle^{(b)}\right],  
\end{eqnarray}
and are given by orbital correlations. Their values at $\eta=0$ are easily 
obtained from the Bethe ansatz result for a 1D Heisenberg chain of 
disordered $a$ and $b$ orbitals (in this case 
$\langle n_{ia}n_{ja}\rangle^{(b)}=\case{1}{4}$): $J_{ab}=0.313$ and 
$J_c=-0.097$. Although the FM interaction along $c$-axis $J_c$ is weaker in 
this limit, it gives a considerable energy gain of $\sim 0.2J$ over the G-AF 
order. It is further enhanced at $J_H>0$ by a mechanism similar to that known 
from cuprates and manganites \cite{Kug82,Ole00}, as the high-spin $^4A_2$ 
state lies by $3J_H\simeq 2.0$ eV below its low-spin $^2E$ counterpart. 
This splitting modifies the 1D orbital-wave (OW) spectrum, 
\begin{equation}
\omega_k^{\rm C}=\sqrt{\Delta^2+R^2(1-\cos^2k)},
\label{gapcaf}
\end{equation}
and the gap $\Delta=\{\eta (R+r)[2R+\eta (R+r)]\}^{1/2}$ opens. Using the 
spin-wave theory we determined the orbital correlations in Eqs. (\ref{jcaf}). 
As a result, one finds increasing (decreasing) FM (AF) exchange constants 
with increasing $J_H$ (Fig. \ref{fig:caf}). Taking a representative value of 
the hopping integral $t=0.2$ eV gives $J=35.6$ meV which leads to the 
exchange constants in the C-phase obtained for a realistic ratio $J_H/U=0.15$: 
$J_{ab}\simeq 7.1$ and $J_c\simeq -9.3$ meV. These values are in the expected 
range, taking the N\'eel temperature $T_N=140$ K of the C-phase in LaVO$_3$. 
We emphasize that the orbital quantum fluctuations play here a dominating role 
and the well known Hund's mechanism due to $J_H$ alone {\it would not 
suffice\/} to obtain $|J_c|>J_{ab}$, giving instead $J_c\simeq -4.4$ meV. 

Next we consider the reasons for the stability of the G-phase in YVO$_3$. 
Unlike LaVO$_3$ with almost equal V--V bonds \cite{Goo95}, this compound 
crystalizes in the distorted structure \cite{Goo95,Kaw94}, indicating that 
the JT effect plays a significant role. It was suggested that energy may be 
gained due to C-type {\it orbital ordering\/}, with $a$ and $b$ orbitals 
staggered in $(a,b)$ planes and repeated along c-axis, while $n_{ic}=1$ 
\cite{Ren00,Miz99,Saw98}. Such ordering can be promoted by
\begin{equation}
H_{\rm JT}=-2V\sum_{\langle ij\rangle\parallel c}\tau_i^z\tau_j^z
            +V\sum_{\langle ij\rangle\parallel (a,b)}\tau_i^z\tau_j^z,
\label{jahnt}
\end{equation}
and {\it competes with the orbital disorder\/}. 
This behavior is remarkably different from the $e_g$ systems, where the JT 
effect and superexchange support each other, inducing 
orbital ordering \cite{Kug82,Tok00,Ole00}. While $V>0$ causes orbital 
splitting by $4V$ and thus lowers the energy of the G-phase ($E_{\rm G}$), 
it has little effect on the energy of the C-phase ($E_{\rm C}$). The energy 
difference is given by
\begin{equation}
\Delta E=E_{\rm C}-E_{\rm G}\simeq V-\case{1}{2}\eta(3R+r)-\delta E_{\rm OW},
\label{deltae}
\end{equation}
where $\delta E_{\rm OW}>0$ is the energy contribution due to quantum 
fluctuations of $t_{2g}$ orbitals. Large spins $S=1$ are almost classical
and their fluctuations could be neglected. 
\begin{figure}[5]
      \epsfxsize=55mm
      \centerline{\epsffile{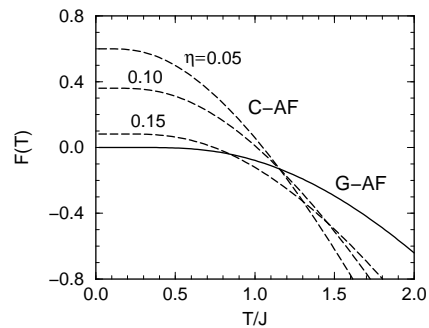}}
\vspace*{0mm}
\caption{
Free energies ${\cal F}(T)$ (in units of $J$) of: 
G-phase obtained with $V=0.65J$ (full line), and C-phase for $\eta=0.05$, 
0.10 and 0.15 (dashed lines), as functions of temperature $T/J$. }
\label{fig:free}
\end{figure}

Orbital excitations are quite different in both AF phases: while the gap 
$\Delta$ is small and grows $\propto\sqrt{\eta}$ in C-phase (\ref{gapcaf}), 
a large gap $\sim 4V$ opens in the OW spectrum of G-phase, 
$\omega_k^{\rm G}=4V+\eta R\cos k$. Thus, both the larger quantum 
fluctuations and additional (classical) energy gain due to finite $J_H$ in 
the C-phase have to be overbalanced by the JT energy $\propto V$ in order 
to stabilize the G-AF order at $T\to 0$. However, the G-phase may be 
destabilized at finite $T$ by larger orbital entropy of the C-phase.
Indeed, taking $V=0.65J$ and $\eta=0.15$, the free energy, 
${\cal F}=\langle {\cal H}\rangle-T{\cal S}$, with the entropy ${\cal S}$ 
determined by orbital excitations, gives a transition from G- to C-phase 
around $T^*\simeq 0.8J$ (Fig. \ref{fig:free}). While this behavior 
reproduces qualitatively the first order transition observed in YVO$_3$ 
\cite{Kaw94,Ren00}, its quantitative description requires a careful 
consideration of lattice and spin entropy contributions to the free energy 
${\cal F}$. These effects are expected to reduce the transition temperature 
$T^*$ down to experimental values.

\begin{figure}[h]
      \epsfxsize=60mm
      \centerline{\epsffile{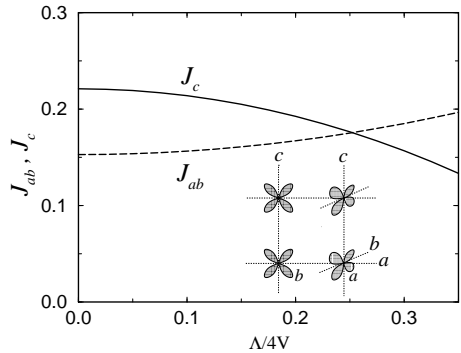}}
\vspace*{3mm}
\caption{
Exchange interactions $J_c$ and $J_{ab}$ (\protect\ref{jgaf}) in G-phase 
(in units of $J$) as functions of spin-orbit coupling $\Lambda/4V$, for 
$\eta=0.15$. The inset shows $a$ and $b$ orbitals, which are staggered in 
$(a,b)$ planes, and repeat themselves along $c$ axis. }
\label{fig:gaf}
\end{figure}

The exchange constants in the G-phase are anisotropic which can be 
understood by analyzing the superexchange expressions (\ref{jcaf}). 
The transitions to the low-spin $^2T_1$ and $^2T_2$ states occur in all 
three directions between pairs of occupied orbitals of the same kind, and 
give the leading AF contribution $\propto (1-\eta r)$. The excitations 
of $d^3\{abc\}$ configurations occur in addition on the bonds in the $(a,b)$ 
planes, and reduce the AF coupling $J_{ab}$ by $\eta R$, giving $J_{ab}<J_c$.
Including in addition a relativistic spin-orbit coupling $\propto\Lambda$ 
($\simeq 18$ meV \cite{Miz96}), we find for the G-phase:
\begin{eqnarray}
\label{jgaf}
J_c&=&\case{1}{4}(1-\eta r)
 -\case{1}{2}(1+2\eta R-\eta r)\bar{\Lambda}^2,  \nonumber \\
J_{ab}&=&\case{1}{4}(1-\eta R-\eta r)
 +\case{1}{4}(1+2\eta R-\eta r)\bar{\Lambda}^2,  
\end{eqnarray}
where $\bar{\Lambda}=\Lambda/4V$. The spin-orbit coupling enhances (reduces) 
the effective superexchange in $(a,b)$ planes (along $c$-axis), as shown in 
Fig. \ref{fig:gaf}. For example, taking a value of $\Lambda/4V=0.2$ which is 
believed to be close to realistic for YVO$_3$ , we find $J_{ab}\simeq 5.9$ 
and $J_c\simeq 6.9$ meV. These values lie in the expected range for the 
G-phase of YVO$_3$. This also demonstrates that the magnetic structure and 
the spin-wave spectrum are completely different depending on the orbital 
state --- the exchange constant $J_c$ which is FM in LaVO$_3$ may become even 
the {\it strongest\/} AF bond when the orbitals have ordered, as in YVO$_3$, 
and the JT splitting (\ref{jahnt}) dominates over the spin-orbit coupling.
 
Summarizing, strong $t_{2g}$ orbital fluctuations in a half-filled system 
of $yz$ and $zx$ orbitals in cubic vanadites lead to a new mechanism of 
ferromagnetic superexchange which stabilizes the C-phase in first 
undistorted LaVO$_3$, and the structural transition follows. The JT effect 
opposes the superexchange and can stabilize the G-phase with orbital ordering 
but only at low temperatures, as the fluctuations of $t_{2g}$ orbitals release 
high entropy, and are thus responsible for the transition from the orbital 
ordered G-phase to the orbital disordered C-phase, observed in YVO$_3$.

We thank B. Keimer, Y. Tokura and L. F. Feiner for stimulating discussions. 
A.M.O. acknowledges the support by the Committee of Scientific Research 
(KBN) of Poland, Project No.~2 P03B 055 20.



\end{multicols} 


\begin{references}

\bibitem{Ima98} M. Imada, A. Fujimori, and Y. Tokura,
                   Rev. Mod. Phys. {\bf 70}, 1039 (1998).

\bibitem{Kug82} K. I. Kugel and D. I. Khomskii,
                   Usp. Fiz. Nauk {\bf 136}, 621 (1982)
                   [Sov. Phys. Usp. {\bf 25}, 231 (1982)].
                   
\bibitem{Tok00} Y. Tokura and N. Nagaosa,
                   Science {\bf 288}, 462 (2000).

\bibitem{Ole00} A. M. Ole\'s, M. Cuoco, and N. B. Perkins,   
                   in: {\it Lectures on the Physics of Highly Correlated 
                   Electron Systems IV\/}, edited by F. Mancini, 
                   AIP Conference Proceedings Vol. {\bf 527} 
                   (New York, 2000).

\bibitem{Fei97} L. F. Feiner, A. M. Ole\'s, and J. Zaanen,
                   \prl {\bf 78}, 2799 (1997).

\bibitem{Kha97} G. Khaliullin and V. Oudovenko, 
                   \prb {\bf 56}, R14\ 243 (1997);
                G. Khaliullin and R. Kilian,
                   J. Phys.: Condens. Matter {\bf 11}, 9757 (1999).   

\bibitem{Kha00} G. Khaliullin and S. Maekawa, 
                   \prl {\bf 85}, 3950 (2000).

\bibitem{Kei00} B. Keimer {\it et al.}, 
                   \prl {\bf 85}, 3946 (2000).

\bibitem{Pen97} H. F. Pen {\it et al.},
                   \prl {\bf 78}, 1323 (1997).

\bibitem{Mil00} F. Mila {\it et al.}, 
                   \prl {\bf 85}, 1714 (2000).

\bibitem{Cas78} C. Castellani, C. R. Natoli, and J. Ranninger,
                   \prb {\bf 18}, 4945 (1978);
                        {\bf 18}, 4967 (1978); 
                        {\bf 18}, 5001 (1978).
                   
\bibitem{Miz96} T. Mizokawa and A. Fujimori,
                   \prb {\bf 54}, 5368 (1996).

\bibitem{Mah92} A. V. Mahajan {\it et al.},
                   \prb {\bf 46}, 10\ 966 (1992).

\bibitem{Goo95} H. C. Nguyen and J. B. Goodenough,
                   \prb {\bf 52}, 324 (1995);
                S. Miyashita, T. Okuda, and Y. Tokura,
                   \prl {\bf 85}, 5388 (2000).

\bibitem{Kaw94} H. Kawano, H. Yoshizawa, and Y. Ueda, 
                   J. Phys. Soc. Jpn. {\bf 63}, 2857 (1994).

\bibitem{Ren00} Y. Ren {\it et al.},
                   \prb {\bf 62}, 6577 (2000);
                M. Noguchi {\it et al.},
                   {\it ibid.\/} {\bf 62}, R9271 (2000).

\bibitem{Miz99} T. Mizokawa, D. I. Khomskii, and G. A. Sawatzky,
                   \prb {\bf 60}, 7309 (1999).

\bibitem{Saw98} H. Sawada {\it et al.},
                   \prb {\bf 53}, 12\ 742 (1996);
                H. Sawada and K. Terakura,   
                   {\it ibid.\/} {\bf 58}, 6831 (1998).

\bibitem{notex} The $t_{2g}^3$ excitation energies expressed in terms of the 
                Racah parameters are: $A-7B$ ($^4A_2$), $A+2B+3C$ ($^2E$ and
                $^2T_1$), and $A+8B+5C$ ($^2T_2$); this spectrum is rigorously 
                reproduced by: $U-3J_H$, $U$, and $U+2J_H$, with $U=A+2B+3C$ 
                and $J_H=3B+C$ respectively [see J. S. Griffith,
                {\it The Theory of Transition Metal Ions}
                (Cambridge University Press, Cambridge, 1971)]. 

\bibitem{Li98}  Y. Q. Li, M. Ma, D. N. Shi, and F. C. Zhang,
                   \prl {\bf 81}, 3527 (1998);
                B. Frischmuth, F. Mila, and M. Troyer,   
                   {\it ibid.\/} {\bf 82},  835 (1999);
                F. Mila, B. Frischmuth, A. Deppeler, and M. Troyer,   
                   {\it ibid.\/} {\bf 82}, 3697 (1999).

\bibitem{Aff87} I. Affleck, T. Kennedy, E. H. Lieb, and H. Tasaki,
                   \prl {\bf 59}, 799 (1987).
                
\bibitem{notet} This is in contrast to the $S=1/2$ case, where 
                (spin-singlet)$\times$(orbital-triplet) and 
                (spin-triplet)$\times$(orbi\-tal-\-singlet) configurations 
                are {\it degenerate\/}, resulting in a strong quantum 
                resonance between them \protect\cite{Kha00}.


\end{references}
\end{document}